\documentclass[12pt]{iopart}
\usepackage{iopams}

\usepackage{graphicx}
\usepackage[utf8]{inputenc}
\usepackage[T1]{fontenc}
\usepackage{dcolumn}
\usepackage{bm}
\usepackage{iopams} 
\usepackage{color}
\usepackage{soul}
\usepackage{url}
\usepackage[normalem]{ulem}
\expandafter\let\csname equation*\endcsname\relax
\expandafter\let\csname endequation*\endcsname\relax
\usepackage{amsmath}

\usepackage{siunitx}
\usepackage{aas_macros}

\usepackage[caption=false]{subfig}
\usepackage{tikz}

\tikzset{%
  base/.style = {rectangle, rounded corners, draw=black,
                  minimum width=0.45\columnwidth, minimum height=3.em,
                  text centered, font=\sffamily},
  inpu/.style = {base, fill=blue!15},
  conv/.style = {base, fill=red!15},
  dens/.style = {base, fill=green!15},
  outp/.style = {base, fill=yellow!15},
}

\begin{document}

\title[Core-Collapse Supernova Search and Deep Learning Classification]{Core-Collapse Supernova Gravitational-Wave Search and Deep Learning Classification}

\author{Alberto Iess}
\ead{alberto.iess@roma2.infn.it}
\address{Universit{\`a} di Roma Tor Vergata, I-00133 Roma, Italy}
\address{INFN, Sezione di Roma Tor Vergata, I-00133 Roma, Italy}

\author{Elena Cuoco}
\address{European Gravitational Observatory (EGO), I-56021 Cascina, Pisa, Italy}
\address{Scuola Normale Superiore, Piazza dei Cavalieri 7, I-56126 Pisa, Italy}

\author{Filip Morawski}
\address{Nicolaus Copernicus Astronomical Center, Polish Academy of Sciences, Bartycka 18, 00-716, Warsaw, Poland}

\author{Jade Powell}
\address{OzGrav, Centre for Astrophysics and Supercomputing, Swinburne University of Technology, Hawthorn, VIC 3122, Australia.}

\vspace{10pt}
\begin{indented}
\item[]December 2019
\end{indented}


\begin{abstract}

We describe a search and classification procedure for gravitational waves emitted by core-collapse supernova (CCSN) explosions, using a convolutional neural network (CNN) combined with an event trigger generator known as Wavelet Detection Filter (WDF). We employ both a 1-D CNN search using time series gravitational-wave data as input, and a 2-D CNN search with time-frequency representation of the data as input. To test the accuracies of our 1-D and 2-D CNN classification, we add CCSN waveforms from the most recent hydrodynamical simulations of neutrino-driven core-collapse to simulated Gaussian colored noise with the Virgo interferometer and the planned Einstein Telescope sensitivity curve. We find classification accuracies, for a single detector, of over 
         $\sim95\%$ for both 1-D and 2-D CNN pipelines. For the first time in machine learning CCSN studies, we add short duration detector noise transients to our data to test the robustness of our method against false alarms created by detector noise artifacts. Further to this, we show that the CNN can distinguish between different types of CCSN waveform models. 

\end{abstract}



\section{Introduction}
\label{sec:intro}

Following upgrades in recent years, the sensitivity of the Advanced LIGO \cite{2010CQGra..27h4006H, aLIGO} and Advanced Virgo \cite{AdVirgo} gravitational-wave (GW) detectors has greatly increased. This led to the first direct observation of GWs emitted from a binary black hole merger in 2015 \cite{PhysRevLett.116.061102} and the first multimessenger observations of a binary neutron star merger \cite{PhysRevLett.119.161101}. Compact binaries are the most common source for ground based GW detectors \cite{catalog} due to their high GW amplitudes, however as the observing time and sensitivity increases, other lower amplitude sources of GWs may be detected. Core-collapse supernovae (CCSNe) are a promising potential future GW source, which have yet to be observed \cite{2016arXiv160501785A, 2019arXiv190803584T}. Their rates for our Galaxy are estimated to be $\sim1$ per 100 years \cite{1991ARA&A..29..363V, 1993A&A...273..383C, Alexeyev2002}.

Searches for compact binary GW signals use a technique known as matched filtering \cite{2016CQGra..33u5004U}. This involves having a large template bank of waveforms that represent the entire parameter space. Producing such a template bank for CCSN signals is currently not possible as the time series waveforms for CCSNe are stochastic, and there are only $\sim30$ CCSNe GW signals currently available from the most state of the art 3D supernova simulations, which do not cover the entire CCSNe parameter space \cite{2018arXiv181205738P, 2019ApJ...876L...9R, 2019MNRAS.486.2238A, 2017ApJ...851...62K}. Therefore, current searches for CCSN GW signals employ a coherent multi-detector network time-frequency analysis in the wavelet domain with minimal assumptions about the signal morphology \cite{2016PhRvD..93d2004K, 0264-9381-25-11-114029, gossan:16}. 

Although the time series of CCSN waveforms is stochastic, common features have been found in time-frequency representations of CCSN GW signals. This includes low frequency (below $\sim200$\,Hz) GW emission due to oscillations of the shock wave known as the standing accretion shock instability (SASI) \cite{0004-637X-584-2-971, 2006ApJ...642..401B, 2007ApJ...654.1006F}, and g-modes that occur at higher frequencies and increase in frequency over time in the spectrograms \cite{2019ApJ...876L...9R, 2017MNRAS.468.2032A, 2018arXiv181205738P, 2019MNRAS.486.2238A, 2017ApJ...851...62K, 2017arXiv170107325Y}. Incorporating some of our knowledge of the time frequency shape of CCSNe into our search methods may increase the sensitivity of our searches in comparison to search methods that make minimal assumptions on signal morphology.

One way to achieve this may be through the application of machine learning techniques. Machine learning techniques have already been developed for compact binary searches \cite{2019PhRvD.100f3015G, 2018PhRvD..97d4039G, PhysRevLett.120.141103}. These searches use a template bank of waveforms for training similar to those used in a matched filter approach. Other studies have applied machine learning techniques to the reduction of short duration transient detector noise artifacts, known as \textit{glitches} \cite{2017CQGra..34c4002P, 2018CQGra..35i5016R, 2017CQGra..34f4003Z}.  For a machine learning CCSN search, a different approach to prepare a training set is needed. Astone et al. \cite{2018PhRvD..98l2002A} apply a convolution neural network (CNN) to search for CCSNe in multiple GW detectors. They use for their training set a phenomenological model that represents g-mode emission in CCSN GW signals. However, they only test their search on one of their phenomenological waveforms and not one of the actual waveforms from 3D hydrodynamical CCSN simulations. In Cavagli\`a et al. \cite{Cavaglia2019}, they perform a single interferometer search for CCSNe by applying machine learning techniques to reduce the single detector noise background. Chan et al., in parallel and
independent to this work, apply a CNN directly to time series data to search for CCSNe. However, they use only the stochastic time series, and therefore do not benefit from the common features expected in the CCSN signals frequency content. They also do not account for the short duration detector noise transients, known as \textit{glitches}, that frequently occur in GW detector data. 

Therefore, in this study we aim to determine the full potential of machine learning for CCSN searches by including spectrograms of CCSNe models, to take full advantage of their common features in frequency space, and to also account for detector glitches. We probe current and future interferometric detector's capabilities of observing GW signals from CCSNe using a combination of an event  trigger generator called Wavelet Detection Filter (WDF) \cite{8553393} and a 1-D or 2-D CNN \cite{Yamashita2018,Russakovsky2015}. 
CNNs are a class of deep neural networks characterised by layers in which the output is computed as a correlation between a filter and the input data to the layer. CNNs have achieved significant performances in image classification tasks and are widely used in the field of computer vision. An in depth description of the subject can be found in Goodfellow et al. \cite{Goodfellow-et-al-2016}. We apply our method to simulated data for the Advanced Virgo detector, as a example of a current ground based detector, and also the future GW detector Einstein Telescope (ET) \cite{Hild_2011}, which has better sensitivity in the CCSN frequency range. To test the robustness of our method, we add glitches to our simulated data and test our algorithm on CCSN waveforms that are not included in the training set.

The paper is structured as follows: Section \ref{sec:waveforms} describes the CCSN waveforms used in this study. Section \ref{sec:dataset}, outlines the method for producing the simulated GW detector noise and glitches. Section \ref{sec:WDF} gives an overview of the WDF trigger generator. Section \ref{sec:ml} describes the machine learning algorithms applied in this study. The results are shown in Section \ref{sec:res}, and a discussion and conclusion is given in Section \ref{sec:disc}.

\section{Supernova Waveforms}
\label{sec:waveforms}

\begin{figure}[!t]
  \centering
    \includegraphics[width=0.48\textwidth]{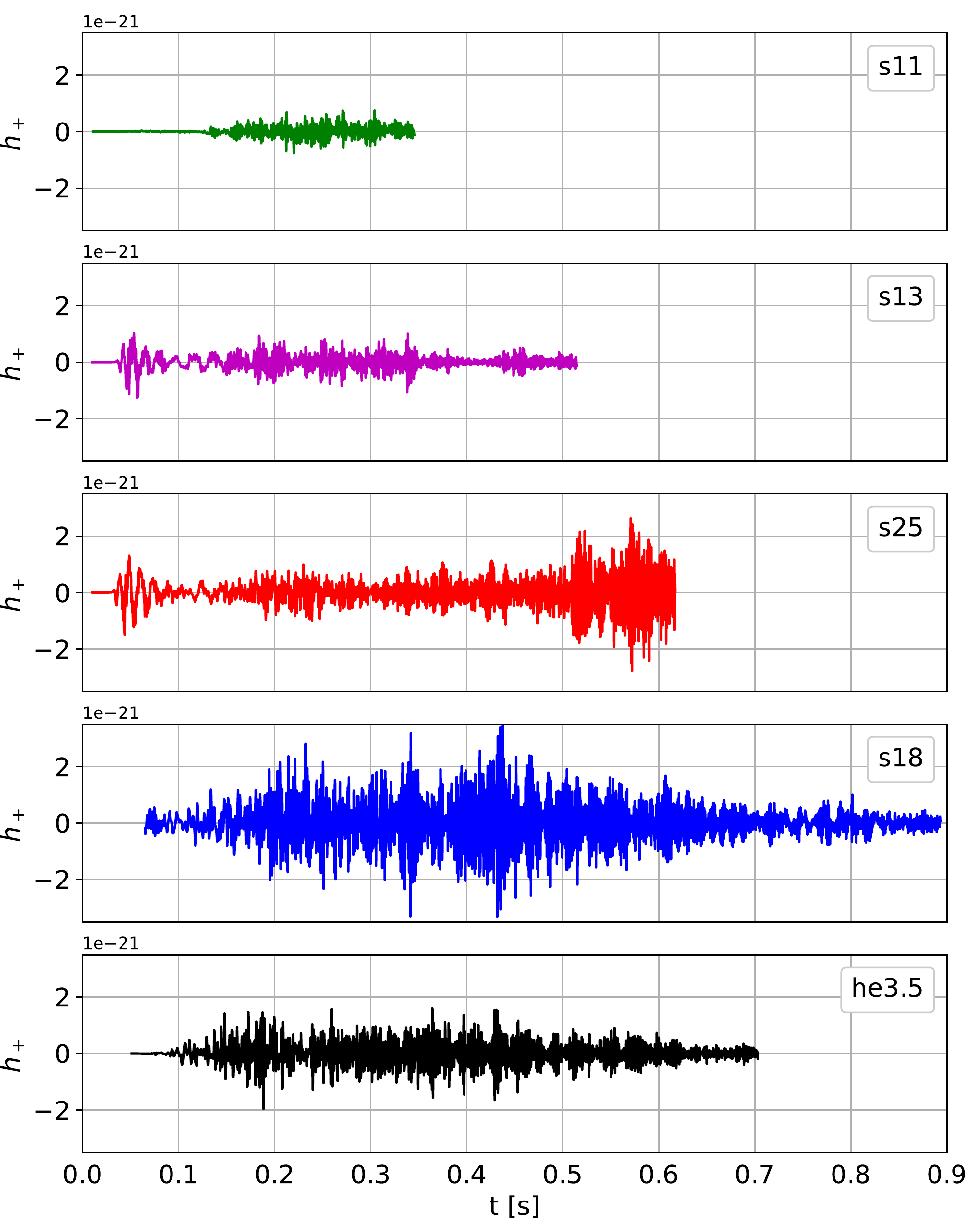}
  \caption{The $\mathrm{h}_+$ gravitational-wave signals for all the CCSN models used in this analysis at a distance of $1$\,kpc, as computed starting from the time of core bounce $t=0$\,s.}
  \label{fig:CCSN_models}
\end{figure}

\begin{figure}[!t]
  \centering
    \includegraphics[width=0.47\textwidth]{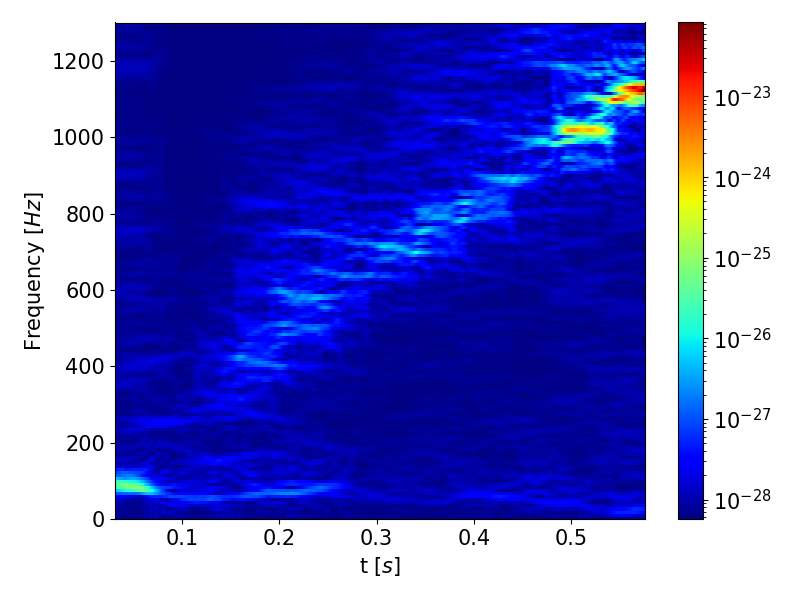}
        \includegraphics[width=0.47\textwidth]{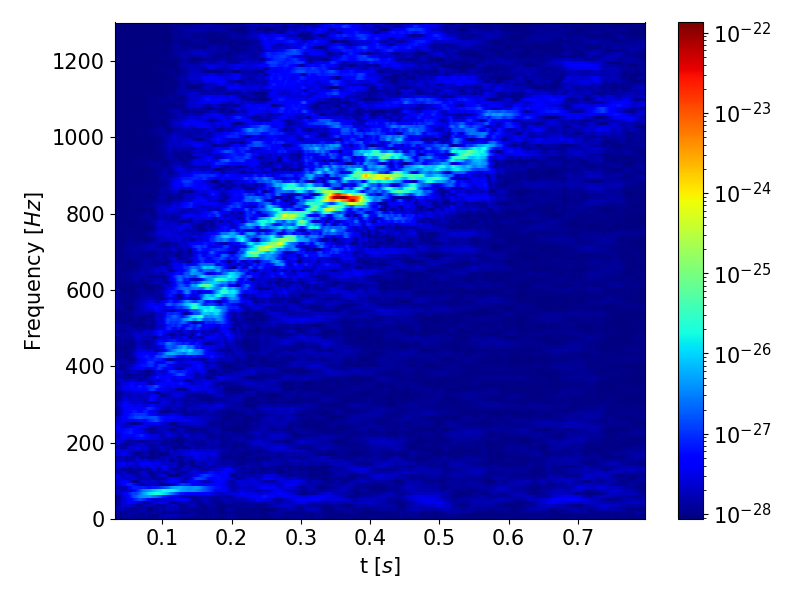}
  \caption{Time-frequency representation of the $\mathrm{h}_+$ polarization of model s25 (\textit{left}) and model s18 (\textit{right}). Both models show high frequency g-modes that peak at different frequencies. Model s25 also has a strong low frequency SASI mode.}
  \label{fig:CCSN_modelsTF}
\end{figure}

In this section, we give a brief description of the CCSN waveforms used in this study. We use the most up to date waveforms available from 3D hydrodynamical simulations of neutrino-driven explosions. Their time series waveforms are shown in Figure \ref{fig:CCSN_models}, while Figure \ref{fig:CCSN_modelsTF} shows examples of their time frequency features.

The neutrino-driven explosion mechanism is thought to occur in most CCSNe (see \cite{2017hsn..book.1095J} for a review). Stars with zero age main sequence (ZAMS) masses above $\sim8\,M_{\odot}$ form electron-degenerate iron cores. When the cores reach their effective Chandrasekhar masses \cite{RevModPhys.62.801}, they become gravitationally unstable and collapse continues until the core reaches nuclear densities. At this stage the core rebounds and a shock wave is launched outwards. The shock wave loses energy as it moves outwards and begins to stall. It must gain more energy for the shock to be revived and power the explosion. In the neutrino-driven mechanism, the energy required to power the explosion comes from a reabsorption of some of the neutrinos. 

\subsection{Model s18}
Model s18 is the $18\,M_{\odot}$ ZAMS progenitor from Powell \& M\"uller \cite{2018arXiv181205738P} simulated with the general relativistic neutrino hydrodynamics code \textsc{CoCoNuT-FMT}. The simulation end time is 0.9\,s, at which time the GW emission has reached very low amplitudes. This model shows a clear g-mode signal in the spectrogram, and reaches amplitudes of $\sim10$\,cm st the source. This model explodes at $\sim300$\,ms after core bounce. The GW frequency peaks at $\sim850$\,Hz. 

\subsection{Model he3.5}
Model he3.5 is the $3.5\,M_{\odot}$ ultra-stripped helium star from Powell \& M\"uller \cite{2018arXiv181205738P} simulated with the general relativistic neutrino hydrodynamics code \textsc{CoCoNuT-FMT}. An ultra-stripped star is a star in a binary system that has been stripped of it's outer layers due to mass transfer to the binary companion star \cite{2015MNRAS.451.2123T}. The simulation ends at 0.7\,s after core-bounce time, well after the peak GW emission phase. This model shows a clear g-mode in the spectrogram. The amplitude of the GW signal is strongest at $\sim900$\,Hz, and reaches an amplitude of $\sim6$\,cm at the source. This model explodes at $\sim0.4$\,s after core bounce. 

\subsection{Model s25}
Model s25 is a $25\,M_{\odot}$ ZAMS model simulated by Radice et al. \cite{2019ApJ...876L...9R} using the Eulerian radiation-hydrodynamics code \textsc{FORNAX}. This model explodes very late at 0.5\,s after core bounce. The simulation ends at 0.62\,s after core bounce time when the GW emission is still high. This progenitor shows a clear signature of the SASI at low-frequency as well as high frequency g-modes. Its GW emission peaks at $\sim1000$\,Hz and reaches amplitudes of $\sim7$\,cm at the source.

\subsection{Model s13}
Model s13 is a $13\,M_{\odot}$ ZAMS model simulated by Radice et al. \cite{2019ApJ...876L...9R} using the Eulerian radiation-hydrodynamics code \textsc{FORNAX}. This model does not explode, and shows GW emission associated with g-modes. This model ends at 0.78\,s after core bounce. Due to the lack of shock revival, this model has lower GW amplitude and peaks at a frequency of $\sim800$\,Hz. This model shows SASI activity at last times, but this does not show as a strong feature in the GW emission. 

\subsection{Model s11}
Model s11 is the $11\,M_{\odot}$ ZAMS model simulated by Andresen et al. \cite{2017MNRAS.468.2032A} with the \textsc{PROMETHEUS-VERTEX} code, which employs Newtonian gravity. The simulation ends 0.35\,s after the core bounce time. This model does not explode. However, it still has a large shock radius which prevents growth of the SASI in this model. This model has the smallest GW amplitude of all the GW signals considered in this study. Its peak frequency occurs at $\sim600$\,Hz. 

\section{Data sets} 
\label{sec:dataset}

\begin{figure}[!t]
  \centering
    \includegraphics[width=0.7\textwidth]{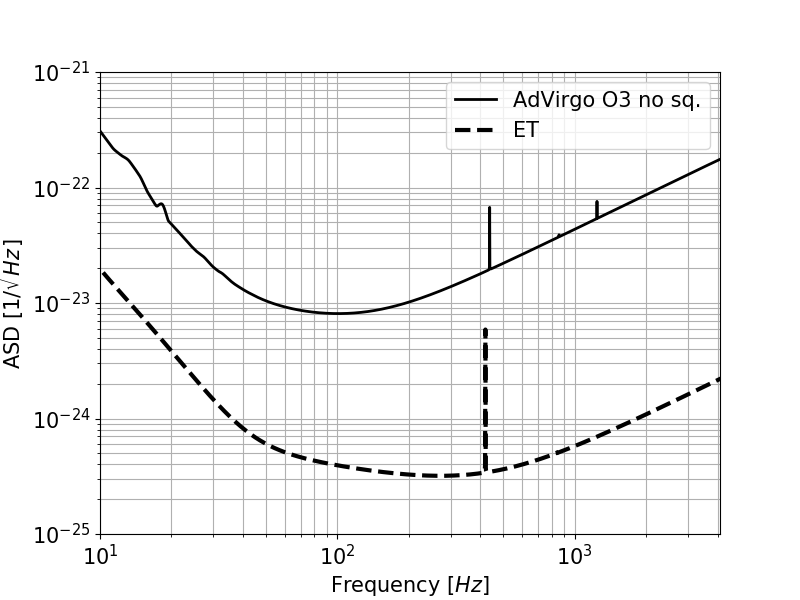}
  \caption{The sensitivity curves of Advanced Virgo at O3 without squeezing, and the Einstein Telescope ET-HF configuration used to produce the noise used in this study. Einstein Telescope is significantly more sensitive at all frequencies where we expect CCSN GW emission.
  }
  \label{fig:sensitivity_curves}
\end{figure}

To test our method, we build different data sets using independent realisations of simulated Gaussian noise for Virgo O3 (VO3) without squeezing \cite{VO3, VO3sensitivity} and the Einstein Telescope (ET) tuned for high frequencies ET-HF from the ET-D configuration described in \cite{Hild_2011}, as examples of current and future GW detectors. The sensitivity curves are shown in Figure \ref{fig:sensitivity_curves}. While ET is designed to be built in a triangle-shaped xylophone configuration, the sensitivities reported in \cite{Hild_2011} refer to a single pair of low (ET-LF) and high-frequency (ET-HF) interferometers of $10$\,km arm length and an opening angle of $90\si{\degree}$. Compared to VO3, ET-HF has improved sensitivity at all frequencies, but especially at the higher frequencies where the peak amplitude of most CCSN models occurs. 

As well as the CCSN waveforms described in Section \ref{sec:waveforms}, in order to test the robustness of our search method we also add to our data ad hoc waveforms representative of noise transients observed in real detector data, glitches, characterised in \cite{detchar1, detchar2}. These include sine Gaussians and waveforms that are a good representation of scattered light glitches, which are a common problem in GW detectors. The simulated glitches are produced in the same method as \cite{cuoco-glitch}, and are given by,

\begin{equation}\label{eq-glitch}
\begin{aligned}[right]
        h_{SG}(t) = h_0 \sin({2 \pi f_0 (t-t_0)}) e^{- \frac{(t-t_0)^2}{2\tau^2}}\\
        h_{SL}(t) = h_0 \sin({\phi_{SL}}) e^{- \frac{(t-t_0)^2}{2\tau}},
\end{aligned}
\end{equation}
where $\phi_{SL} = 2\pi f_0 (t-t_0)[1-K(t-t_0)^2]$, and $\tau$ is the time parameter which describes signal width. The parameter K defines the curvature of the scattered light arches in the time-frequency domain. We used values for $\tau$ and $K$ such that the glitches are contained in time windows comparable to those of the CCSN signals.  This is compatible with the values in \cite{cuoco-glitch} based on comparison with real detector glitches. We define the quality factor of sine Gaussians as $Q = 2\pi f_0 \tau$ and its values are determined by the central glitch frequency $f_0$ and the exponential decay time constant $\tau$. In interferometric data, glitches which can be modelled by sine Gaussians generally have  $Q<100$. We also included glitches with $Q>100$ so that they will have longer durations similar to the CCSNe models. The central frequency range varies from $30$ to $1000$\,Hz. The second equation in Eq.(\ref{eq-glitch}) is representative of scattered light glitches and their harmonics at lower frequencies and values of $K\sim0.5$. 
The equation for $h_{SL}$ can also describe a type of glitch caused by radio frequency beat notes, known as whistle glitches, which occur at higher frequencies and $K$.   
We picked the amplitude value $h_0$ in terms of background noise standard deviations from a log-normal distribution with mean and sigma for the underlying normal distribution of $\mu_g = -1$ and $\sigma_g = 2$. This choice results in $h_0$ typically below one noise standard deviation and signal-to-noise ratios (SNRs) between 1 and 1000 for both kinds of glitches. However, the peak of the SNR distribution for the scattered light glitches falls at higher values of $\sim50$ than the peak for the sine Gaussian glitches at $\sim30$.

We define the square of the injection SNR in the frequency domain as in \cite{PhysRevD.49.2658},\cite{Maggiore}

\begin{equation}
    \left(\frac{S}{N}\right)^2 = 4  \int_0^{f_{max}}{ \frac{\tilde{h}(f) \tilde{h}(f)^* }{S_n(f)} df},
\end{equation}
where $\tilde{h}(f)$ is the Fourier transform of the injected waveform time-series $h(t)$, the asterisk denotes the complex conjugate and $S_n(f)$ is the one-sided noise power spectral density. 

\begin{figure}[!t]
    \centering
        \includegraphics[width=0.7\textwidth]{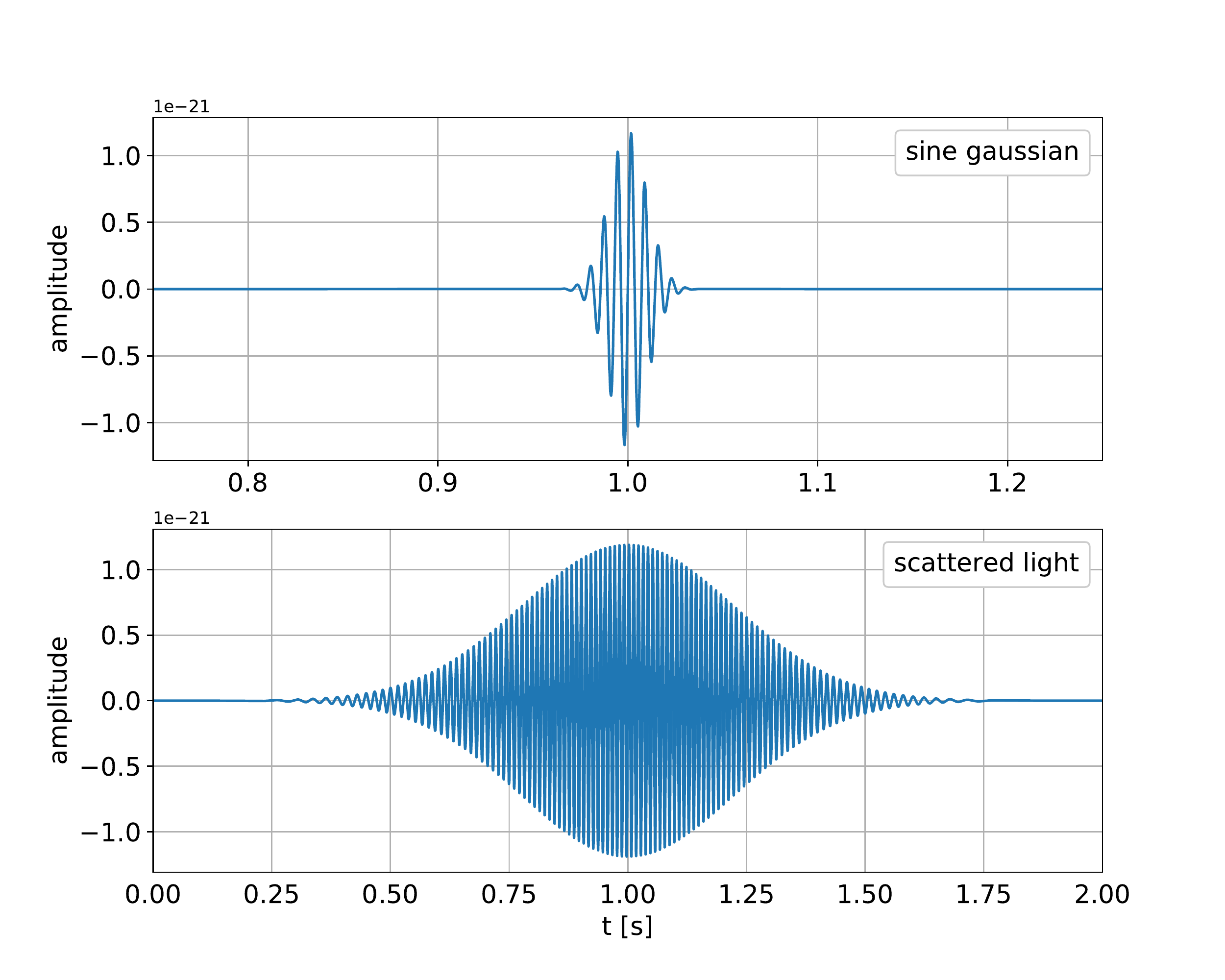}
    \caption{Examples of glitch time-series: a sine gaussian with $Q=8.75$ and a scattered light glitch.}
    \label{fig:binary_classification_VO3_all}
\end{figure}


For each interferometer, we produced a total of 5 hours of detector noise over which we injected signals at a rate of 1000 per hour for each model, including the two noise glitch models. The simulated CCSN sources follow uniform distributions in sky angles and the distances $r$ are log-normally distributed in order to cover different order of magnitudes and SNRs. Detector sensitivity curves influence our choice for the distance distributions used in  the two datasets: for VO3 distances range from 0.01\,kpc to 10\,kpc  with average $\bar{r} \approx 0.6$\,kpc, while for ET the distances range from 0.1\,kpc to 1000\,kpc with $\bar{r} \approx 70$\,kpc.
The angles are taken in the coordinate system described by the interferometer, so that the signal waveform is given by the weighted sum of the two GW polarizations $h_+$ and $h_\times$ as

\begin{equation}
    h(t) = F_+ h_+(t) + F_\times h_\times(t),
\end{equation}
where $F_+$ and $F_\times$ are the antenna pattern functions given by

\begin{equation}
\begin{aligned}
    F_+ = \frac{1}{2} (1+\cos^2{\theta}) \cos{2\phi}\cos{2\psi} - \cos{\theta} \sin{2\phi} \sin{2\psi}\\
    F_\times = \frac{1}{2} (1+\cos^2{\theta}) \cos{2\phi}\sin{2\psi} + \cos{\theta} \sin{2\phi} \cos{2\psi}, 
\end{aligned}
\end{equation}
where $\phi$, $\theta$ are the angles to the source measured from one of the interferometer arms and from the normal to the detector plane, respectively,  and $\psi$ is the polarization angle \cite{schutz2011}. The use of the ET-HF noise curve allows us to adopt the single interferometer antenna pattern as an approximation also for ET. We will drop the subscript and refer to ET-HF as ET in the rest of the text.
The waveforms are resampled to 4096\,Hz, as most of their GW emission is below 1000\,Hz. 

\section{Wavelet Detection Filter}
\label{sec:WDF}

Trigger event generation is provided by Wavelet Detection Filter (WDF) \cite{8553393}, which has been previously used in the context of glitch classification problems as described in Powell et al. \cite{2015CQGra..32u5012P}. 
The detection algorithm is based on decomposition of the data into multiple time-frequency resolution maps, through the wavelet transform. A generic time series $s(t)$ is projected onto a family of mutually orthonormal wavelets as

\begin{equation}
    \langle s | \psi_{a,b} \rangle = \int_{-\infty}^{+\infty} s(t) \frac{1}{\sqrt{b}} \psi^* \left( \frac{t-a}{b} \right) \, dt ,
\end{equation}
where $\psi^*$ is the complex conjugate of the mother wavelet, parameter $b$ sets the scale of the time-frequency map and $a$ is the time-shifting parameter.  For analysis of transients WDF implements the discrete wavelet transform using a bank or wavelet family including  Daubechies, Haar and spline wavelets~\cite{daubechies1992ten, Mallat, Unser1997}.  

In this study, we whiten the data in the time domain as described in Cuoco et al. \cite{cuoco-whitening}.
This is achieved by fitting the noise power spectral density (PSD) using an Auto-Regressive (AR) model as described in \cite{cuoco-whitening}.
In this study, the initial 300 seconds of each dataset are used to fit the  noise PSD with an AR model of order 4000. After whitening, the wavelet coefficients are expected to contain features of the transient waveforms on different time and frequency scales. In particular, only relevant coefficients are retained using the Donoho and Johnstone method \cite{10.1093/biomet/81.3.425} which sets a lower threshold $t$ on the absolute value of the wavelet coefficients $|w_i|$ as

\begin{equation}
    t = \sqrt{2\log{N}} \hat{\sigma},
\end{equation}
where $N$ is the number of window data points and $\hat{\sigma}$ the estimation of the noise standard deviation. WDF describes the event triggers by means of a set of  parameters: the trigger time (timestamp), the mean and maximum frequency content of the event, the wavelet coefficients containing the event information and a value proportional to the WDF $\mathrm{SNR}_{w}$, not to be confused with the injection SNR, given by

\begin{equation}
    \mathrm{SNR}_{w} = \frac{\sum_i w_i^2}{\hat{\sigma}},
\end{equation}
which is the indicator for the threshold of the WDF detected events.

The WDF window size used in this analysis is 1024 points with an overlap of 256 between consecutive windows, which corresponds to a time-window of 0.25\,s and 0.0625\,s respectively for a sampling frequency of 4096\,Hz.

After WDF finds all the triggers in the data, the time-domain whitened strain data around the triggers are then fed to the CNN classifier. The advantages of combining a trigger generator with the CNN are that the trigger generator saves time by not needing to apply the CNN to all data, and as the CNN only produces a classification of events, information provided by the event trigger generator, such as the SNR, is required to produce an estimate of the significance of a real detection. 

\section{Deep Learning pipelines}
\label{sec:ml}

Each candidate event found by WDF \cite{8553393} has a timestamp which is used to build the input sample data for the CNN \cite{Russakovsky2015, Yamashita2018}. To mimic the real case scenario, among all trigger times we choose the one corresponding to the highest $\mathrm{SNR}_w$ value inside a time window coincident with the injection. We define the coincident window so that the WDF window overlaps the true signal by at least 3/4 of its length (i.e. 768 points). Note that we are assuming that the trigger timestamp produced by WDF occurs approximately at half of the injected signal duration, so that we can take a symmetric $1$\,s time window around it to cover the entire waveform.  To take into account the fact that the peak  $\mathrm{SNR}_w$ may not occur at such exact time, a random shift in time can be manually applied, although this has not been done to achieve the results presented in this analysis. Given the timestamp, the dataset can be tailored for the two cases of 1-D and 2-D convolutional neural networks. The full dataset is divided in the following way: $60\%$ for model training, $10\% $ for model validation and $30\%$ for testing.

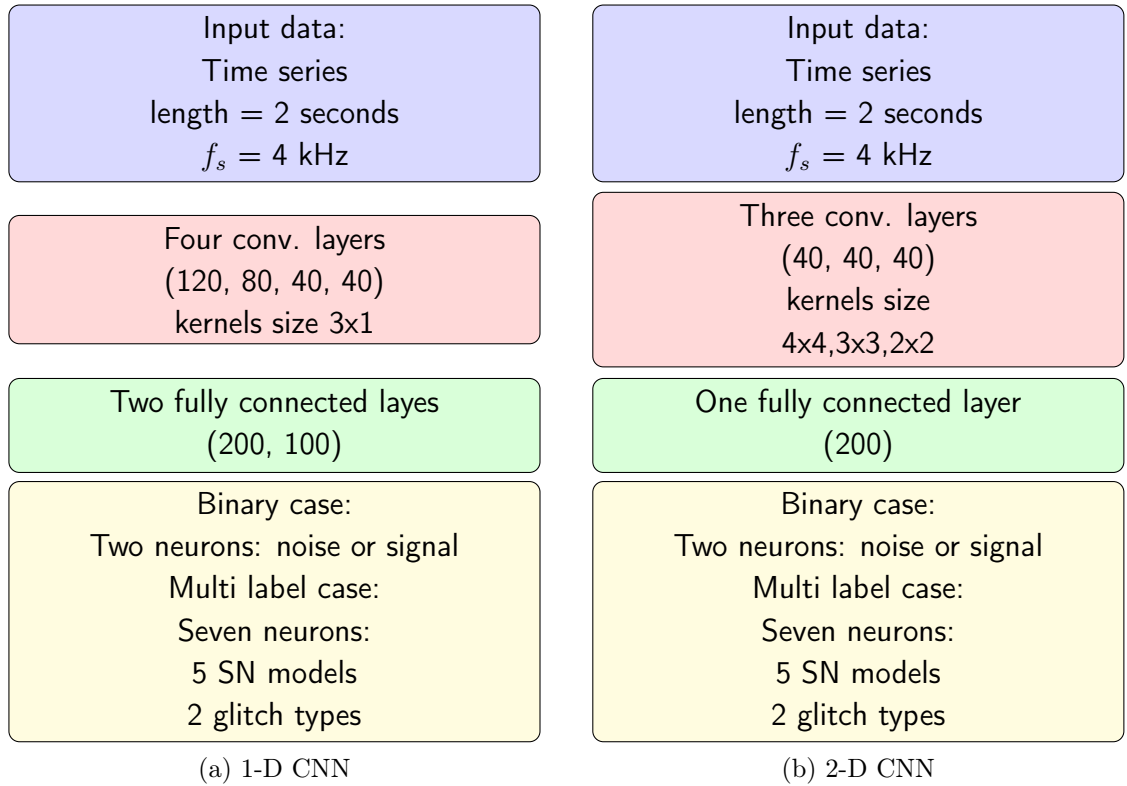
\begin{figure}[htbp]
\subfloat[1-D CNN]{{
\begin{tikzpicture}[node distance=1.5em, 
    every node/.style={fill=white, font=\sffamily}, align=center]
  \node (inpu1)  [inpu]                                 {Input data:\\Time series\\length = 2 seconds\\$f_s$ = 4 kHz};
  \node (conv1)  [conv, below of=inpu1, yshift=-4.5em]    {Four conv. layers\\(120, 80, 40, 40)\\kernels size 3x1};
  \node (dens1)  [dens, below of=conv1, yshift=-3.2em]  {Two fully connected layes\\(200, 100)};
  \node (outp1)  [outp, below of=dens1, yshift=-4.5em]  {Binary case:\\Two neurons: noise or signal\\
  Multi label case:\\Seven neurons:\\
  5 SN models\\2 glitch types};
  \label{fig:arch1}
\end{tikzpicture}
}}
\quad 
\subfloat[2-D CNN]{{
\begin{tikzpicture}[node distance=1.5em,
    every node/.style={fill=white, font=\sffamily}, align=center]
  \node (inpu1)  [inpu]                                 {Input data:\\Time series\\length = 2 seconds\\$f_s$ = 4 kHz};
  \node (conv1)  [conv, below of=inpu1, yshift=-4.5em]    {Three conv. layers\\(40, 40, 40)\\kernels size\\4x4,3x3,2x2};
  \node (dens1)  [dens, below of=conv1, yshift=-3.2em]  {One fully connected layer\\(200)};
  \node (outp1)  [outp, below of=dens1, yshift=-4.5em]  {Binary case:\\Two neurons: noise or signal\\
  Multi label case:\\Seven neurons:\\
  5 SN models\\2 glitch types};
  \label{fig:arch2} 
\end{tikzpicture}
}}
\caption{An overview of the two searches used in this study. (a) describes the steps of our 1-D CNN search and (b) describes the steps of the 2-D CNN search.}
\end{figure} 

\subsection{CNN 1-D}

The input data consist of 1\,s of whitened time-series sampled at 4096\,Hz, symmetrically taken around the event trigger times. Longer segments lead to a decrease in accuracy, since we are inserting more noise background before and after the signals. The network is structured in four layers as showed in Figure~\ref{fig:arch1}: each of them contains a convolutional layer with respecively 120, 80, 40 and 40 filters with 3x1 kernels and a rectified linear unit activation function (we recall that $ \operatorname{ReLU}(x) = max(0,x)$). A 2x2 max pooling is then applied to reduce the input data to the next layer. Spatial dropout involving $40\%$ of the layer's weights is applied as a regularization procedure. Two fully connected (FC) layers with 200 and 100 neurons follows. The final layer depends on the task given to the CNN. In the case of binary classification between signal and noise instances (or two different kinds of signal) there will be a $2$-neuron FC layer with softmax activation function and the associated loss function will be the binary cross-entropy. In the case of multilabel classification this will be substituted by a $n_{l}$-neuron FC layer, where $n_{l}$ is the number of distinct labels, and a categorical cross-entropy loss function. The optimizer used is Adam \cite{2014arXiv1412.6980K}, with a learning rate $\alpha=0.001$ and a batch size of 128 samples per training step. The number of epochs is never taken greater than 50, since at larger values the model starts to overfit.

\subsection{CNN 2-D}
In the 2-D case, inputs are spectrogram images built using a total of 2\,s whitened data around the trigger times, although the window is tunable for shorter signals. By using spectrograms, information on the signal phase, which is dependent on the stochastic realization provided by simulations, is lost and replaced by the added value of the frequency morphology of the waveform.  The architecture is similar to the 1-D case, but with some relevant differences as sketched in Figure~\ref{fig:arch2}, starting from three layers instead of four. The number of convolutional filters in each layer is 40. Moreover, the kernels are taken to have different sizes (4x4, 3x3, 2x2) and no spatial dropout is involved. Finally the batch size is of 64 samples while the Adam learning rate is  $\alpha = 0.001$, with an associated learning decay rate of 0.066667.

\section{Search and Classification Results} 
\label{sec:res}

\begin{figure}[htb]
  \centering
    \includegraphics[width=0.75\textwidth]{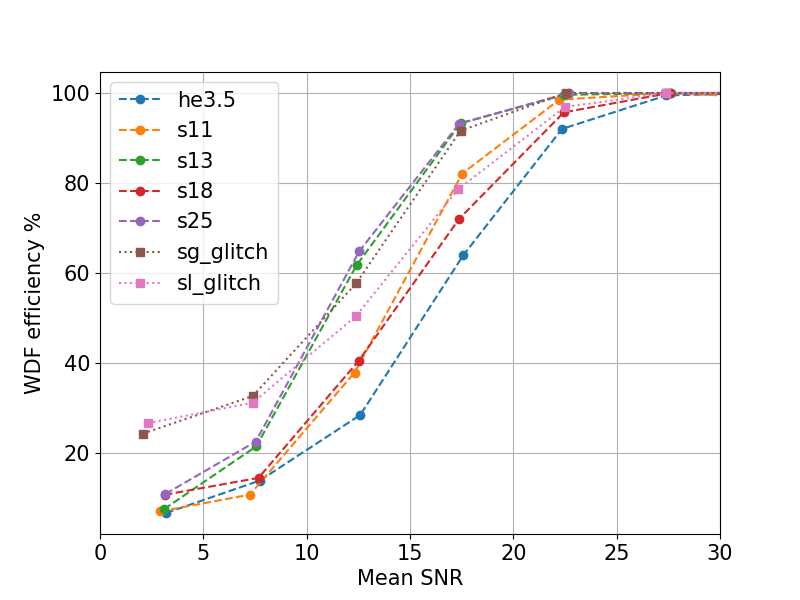}
  \caption{WDF signal trigger detection efficiency as a function of the average injection SNR for supernova signals and glitches in the VO3 data. Results are similar for signals in ET data. At SNR$>30$ the efficiency is $100\%$.}
  \label{fig:WDF_VO3_efficiency}
\end{figure}

First we apply WDF to the data to find the trigger times at which the signals and glitches occur. Figure \ref{fig:WDF_VO3_efficiency} shows the WDF detection efficiency as a function of the average injection SNR for the VO3 data, computed on segments each containing 3000 signals of a specific class. All signals with $\mathrm{SNR}>25$ are detected by WDF. 
The SNR distribution depends on two factors: distance and model GW amplitude. We use the same distance range for all CCSN models over the same detector background, while there is a lot of variation in the models amplitudes (as previously shown in Figure \ref{fig:CCSN_models}). At fixed distances the low energy, non exploding models, such as model s11, have lower SNRs and therefore the SNR distributions are not the same for each models training and testing set.
Models s13 and s25 have the highest WDF detection efficiency, which is due to their models having higher SNR waveforms in the training data. Model s18 is also a high amplitude model, but its detection efficiency may be worse than the other high amplitude models due to its longer duration. We see similar WDF results for the signals and glitches in ET noise. 

\subsection{Supernova vs Glitch Classification}

In this section, we perform a binary classification into two categories: signals and glitches. The signal class should include all CCSN models, and the same holds for the glitch class. In a real case scenario, training would be carried out on all available realistic CCSN waveforms. The true positive rate (TPR)
for each class are found on the diagonal of the confusion matrix and are defined in the test set as the ratio between the correctly classified samples, true positives, and the total number of samples in a class computed as the sum of true positives (TP) and false negatives (FN) as 

\begin{equation}
    TPR = \frac{TP}{TP+FN}.
\end{equation}{} 

\subsubsection{Models included in training}

First we test our method using the same waveforms in the test sets as we used in the training. This is not a very realistic case, as we know that a real CCSN detection is unlikely to look exactly like one of our training waveforms. However, it does allow us to determine the maximum accuracy of our method, and we can compare that number to the case where different waveforms are used for training and testing to quantify how much accuracy is lost due to a lack of CCSNe waveforms. 

\begin{figure}[htb]
    \centering
        \includegraphics[width=0.45\textwidth]{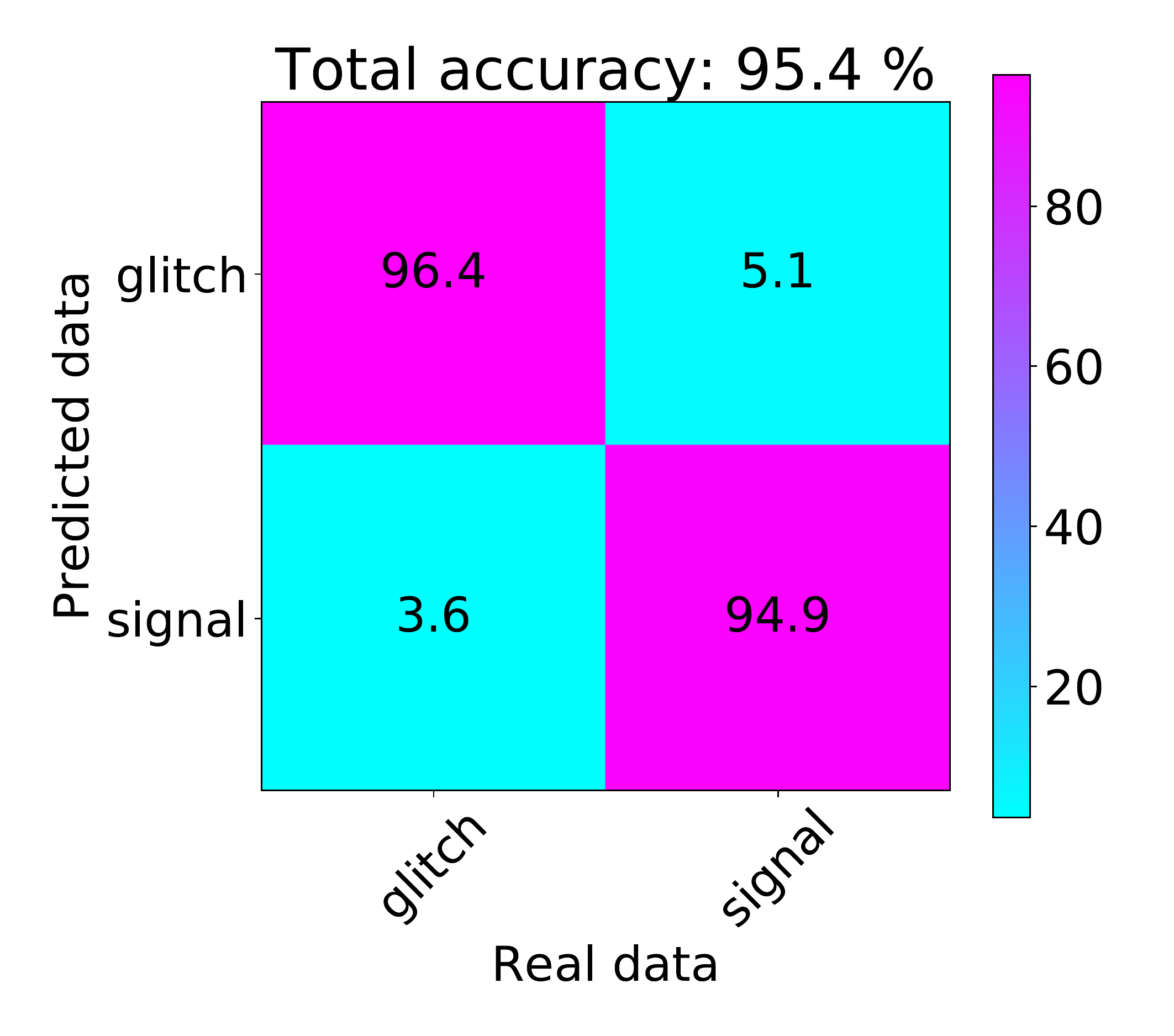}
    \centering
        \includegraphics[width=0.45\textwidth]{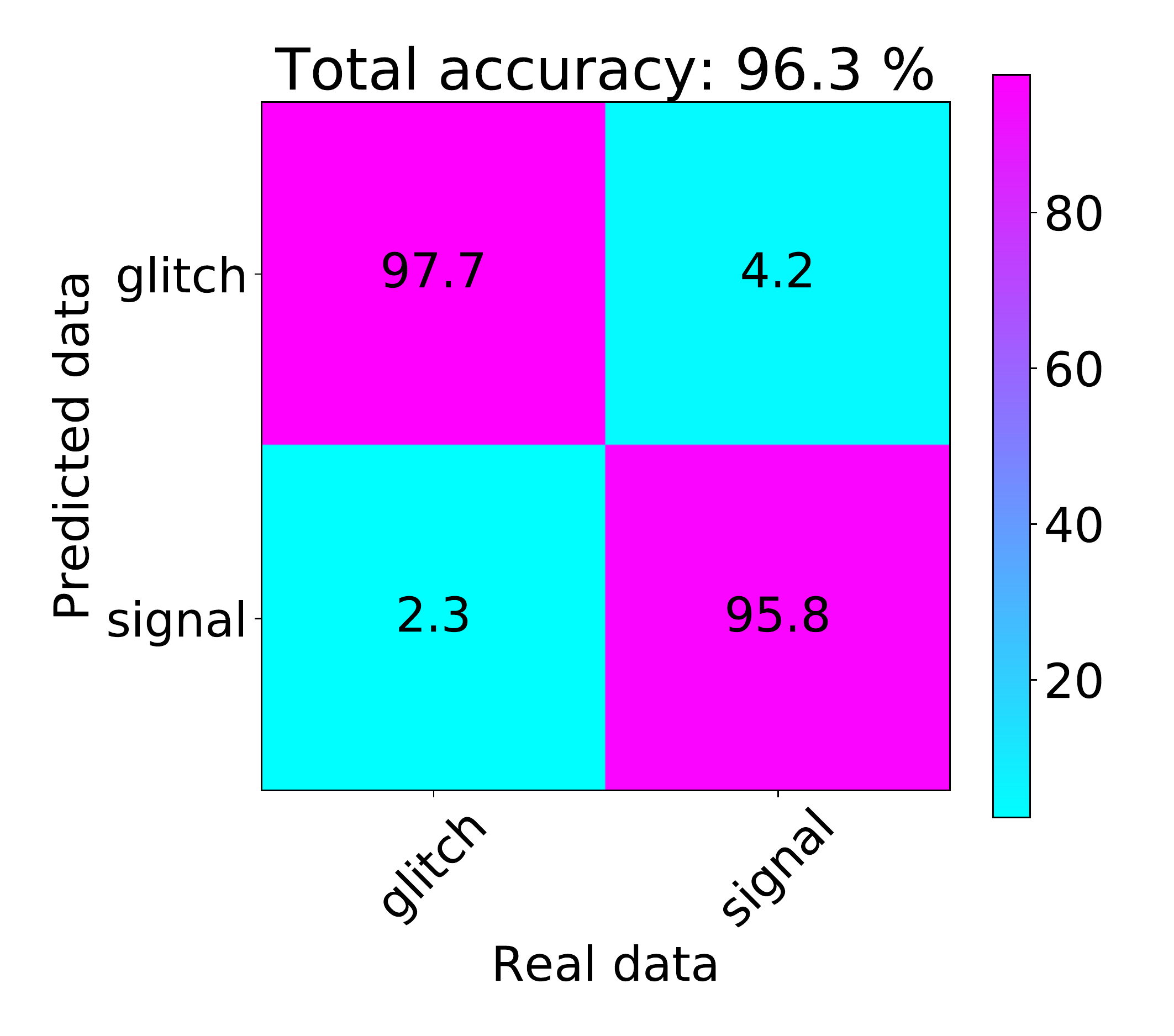}
    \centering
        \includegraphics[width=0.45\textwidth]{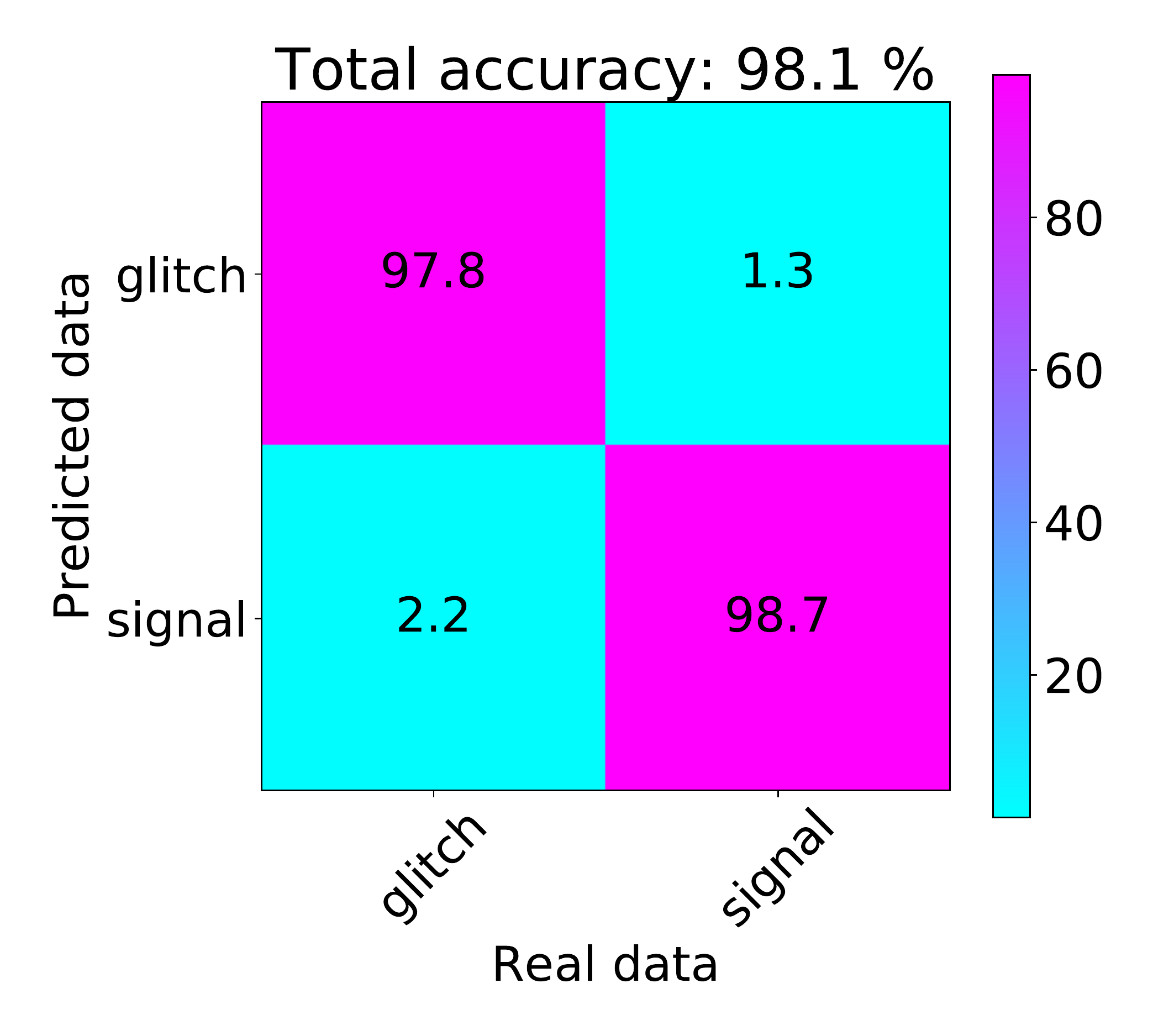}
    \centering
        \includegraphics[width=0.45\textwidth]{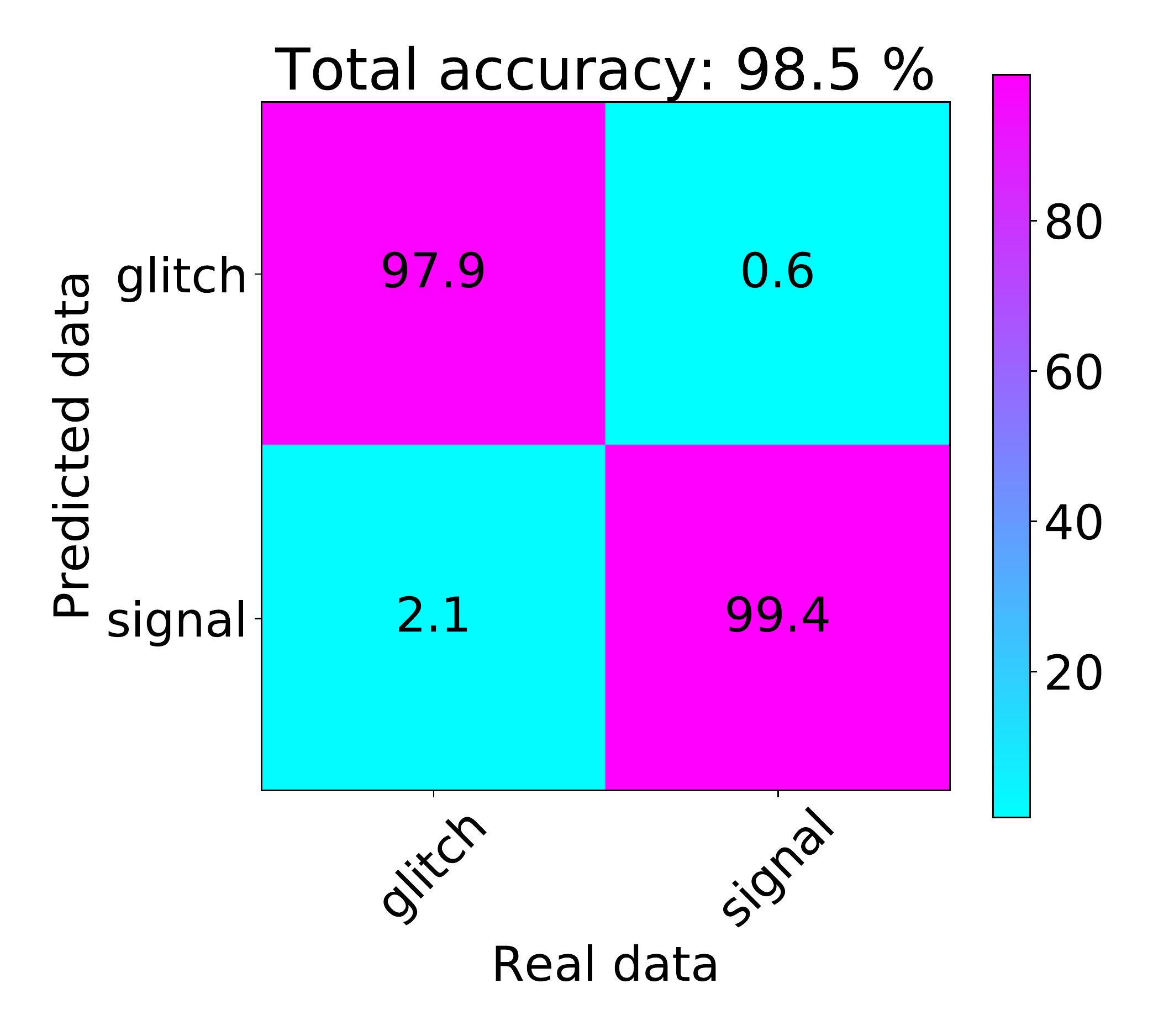}
    \caption{Binary 1-D CNN (\textit{left}) and 2-D CNN (\textit{right}) classification of glitches and CCSN signals from all considered models, added to simulated Virgo O3 (\textit{top}) and Einstein Telescope (\textit{bottom}) noise background.}
    \label{fig:binary_classification_VO3_all}
\end{figure}

The confusion matrix for this case, using the VO3 data, is presented in Figure \ref{fig:binary_classification_VO3_all}. We find very high accuracies (above $\sim95\%$) for both the 1-D and 2-D CNN results. The results improve slightly for the 2-D case as expected due to the common features in the CCSN spectrograms. 

The misclassified waveforms correspond to low SNRs, indistinctive of the CCSN model. For the ET dataset, results are better both in the 1-D the 2-D case with accuracies in the signal class  $\sim99\%$, with $\sim98\%$ of the total samples correctly classified when including glitches, as showed in Figure~\ref{fig:binary_classification_VO3_all} (\textit{bottom}). The improved results for ET are due to the better high frequency sensitivity of the detector.

\subsubsection{Models not included in training}
As a real CCSN signal would not match exactly the waveforms we use for training, we test the ability of the CNN to generalize to different CCSN models by training the network on a subset of four signal models and then testing it on the one that was not included in the training step. 

The sensitivities achieved for the signal classes are reported in Table \ref{tab:accuracies} for the two detectors, VO3 and ET, and the two CNN implementations, along with the total accuracies. The \textit{Total} accuracy is defined as the overall number of correctly classified samples in either the signal or noise class, over the total number of samples. The results are similar for both the 1-D and 2-D CNNs with total accuracies consistently above $90\%$. The accuracy is not as high for model s25 as it is for the other models. We suspect that this is due to model s25 being the only model that we consider that contains a strong low frequency SASI mode. Model s25 also has a higher peak GW frequency than the other models used for training. Therefore, we could improve on this result in the future by including more CCSN waveforms with SASI features in the training set. The results show that we do not lose a lot of accuracy when we perform a more realistic test with models not included in the training set.  

\begin{table}[h!]
  \begin{center}
    \begin{tabular}{|c|c|c|c|c|}
      \hline
      {\textbf{VO3}} &
      \multicolumn{2}{c} {\textbf{1-D CNN accuracy}} & \multicolumn{2}{|c|} {\textbf{2-D CNN accuracy}}\\
      \hline
      \textbf{Test set} & \textit{\, Signal \, } & \textit{Total} & \textit{\, Signal \, } & \textit{Total}\\
      \hline
      s11 & $93.9$ & $93.7$ & $98.0$ & $94.3$ \\
      \hline
      he3.5 & $96.2$ & $95.5$ & $95.2$ & $97.6$ \\
      \hline
      s18 & $97.5$ & $96.7$ & $98.4$ & $97.9$ \\
      \hline
      s13 & $94.5$ & $94.4$ & $94.4$ & $96.9$ \\
      \hline
      s25 & $95.1$ & $95.1$ & $92.2$ & $95.9$ \\
      \hline
    \end{tabular}
  \end{center}
  \begin{center}
    \begin{tabular}{|c|c|c|c|c|}
      \hline
      {\textbf{ET}} &
      \multicolumn{2}{c} {\textbf{1-D CNN accuracy}} & \multicolumn{2}{|c|} {\textbf{2-D CNN accuracy}}\\
      \hline
      \textbf{Test set} & \textit{\, Signal \, } & \textit{Total} & \textit{\, Signal \, } & \textit{Total}\\
      \hline
      s11 & $94.5$ & $96.7$ & $95.5$ & $97.2$ \\
      \hline
      he3.5 & $98.0$ & $97.8$ & $98.5$ & $97.6$ \\
      \hline
      s18 & $92.1$ & $94.2$ & $92.4$ & $96.2$ \\
      \hline
      s13 & $95.9$ & $96.6$ & $84.5$ & $94.1$ \\
      \hline
      s25 & $73.3$ & $83.2$ & $89.6$ & $95.5$ \\
      \hline
    \end{tabular}
    \caption{Binary classification results on CCSN models not included in training. For the 1-D and 2-D CNN implementations and both detector datasets, we show the sensitivity for the \textit{Signal} class. We also report the \textit{Total} accuracy computed as the ratio of the number of correctly classified samples in one of the two classes and the total number of test samples. The waveform used in the test set for the signal class is shown in the first column.}
    \label{tab:accuracies}
  \end{center}
\end{table}

\subsection{Multi-model Classification}

In this section, we aim to distinguish between different CCSN waveform models. Distinguishing between different types of CCSN waveforms can allow us to determine properties of a CCSN GW source, such as the explosion mechanism, or the presence of features like the low frequency SASI \cite{2016PhRvD..94l3012P, 2019PhRvD..99f3018R}. The models we use in this section all have the same neutrino-driven explosion mechanism, and are used just to demonstrate the method rather than to make any astrophysical statement by distinguishing between them with the CNN. In the future, we could apply the same technique to models with different explosion mechanisms when more modern 3D CCSN waveforms, for example for the magnetorotational explosion mechanism, become available. 

CCSN waveform end times are usually determined by a lack of available computer time rather than some astrophysical reason. Therefore, when we attempt to distinguish between them we do not want the CNN to include information about the different lengths of the signals, as that information is not astrophysical. To avoid this issue, in the 2-D CNN case we adjusted the window size to match the length of the shortest waveform considered in this analysis, s11, with duration $0.35$\,s. 

The morphology of the time series waveforms are stochastic, and therefore cannot be related back to individual types of models. However, it is expected that their time series amplitudes may be related to the CCSN properties. In this section, to avoid problems related to the stochastic nature of the waveforms, we decide to ensemble~\cite{ref10.1007/3-540-45014-9_1} the results for both the 1-D and 2-D CNN to produce the multi-label classification results. 

\begin{figure}[t]
  \centering
    \includegraphics[width=0.6\textwidth]{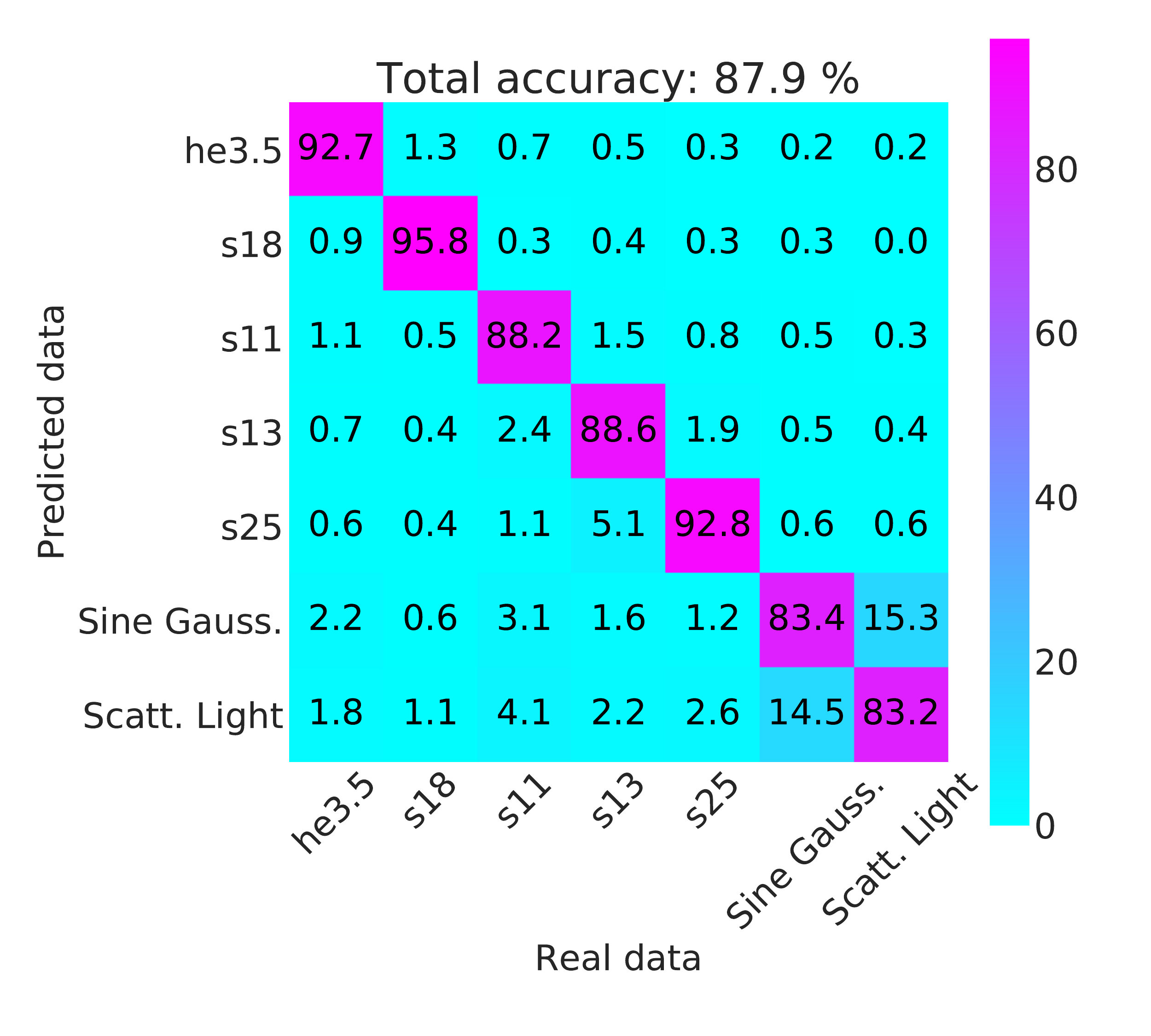}
  \centering
    \includegraphics[width=0.6\textwidth]{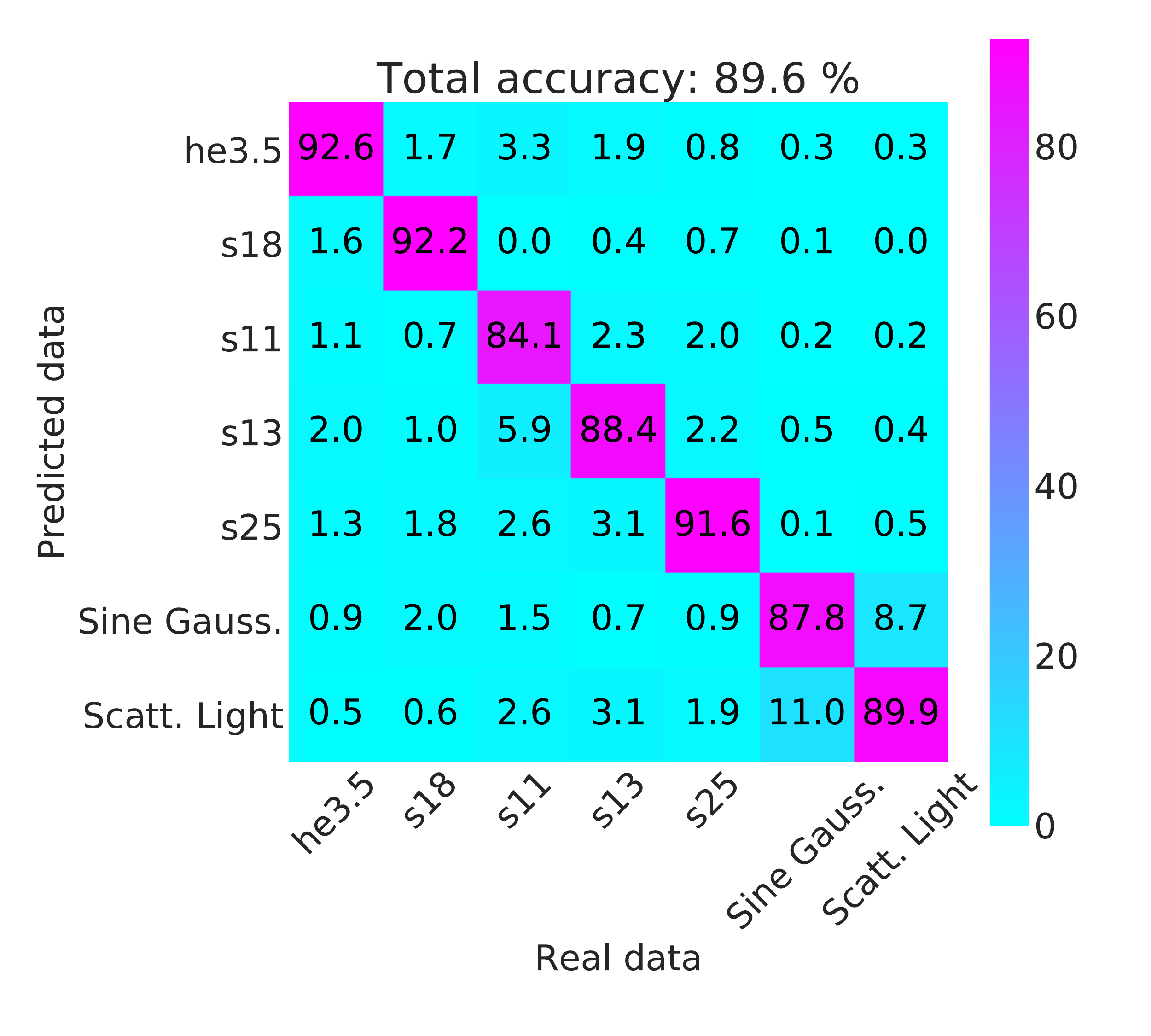}
  \caption{Confusion matrices for multilabel classification for the VO3 (\textit{top}) and ET (\textit{bottom}) datasets with all the CCSNe and the two glitch models.}
  \label{fig:merged_multilabel}
\end{figure}

The confusion matrices for the ensemble multilabel classification are shown in Figure  \ref{fig:merged_multilabel}. The sensitivities for each class are found on the diagonal. While total accuracies decrease compared to the binary problem, they still reach $87.9\%$ for VO3 and $89.6\%$ for ET. In both cases, higher emission models are classified more accurately. Sine Gaussians with longer duration $\tau$ and shorter scattered light glitches are responsible for a large part of the misclassified samples among the two noise classes. The trained model is able to distinguish among the different CCSN models even when using a short time window to build the sample spectrogram and time-series.

\section{Discussion} 
\label{sec:disc}

Current searches for GWs from CCSNe make no assumptions on the signal morphology. The time series waveforms for CCSN GW signals are stochastic, and so they can never be used in a matched filter search. However, the frequency content of CCSN signal predictions is not stochastic and can be directly related to the properties of the explosion and the proto neutron star. A matched filter search in the frequency domain still cannot be performed for CCSN signals because there are not enough 3D waveforms currently available, and they do not fully cover the CCSN parameter space. Further to this, many simulations are ended before the peak GW emission time or have missing input physics. However, in the waveforms we do have available, we see similar features in their spectrograms that we would expect to see in a real GW detection, and incorporating this information into our GW searches may increase our sensitivity to these sources.  

We aimed to achieve this by performing a new search using a combination of a trigger generator called WDF and a CNN. A trigger generator finds all excess power events above a threshold, and provides some information about each event. The events could be real GW signals or background detector noise glitches. Using a trigger generator allowed us to save search time because we then only feed time around the WDF triggers to the CNN. Using a trigger generator also enables us to estimate how significant each GW detection is, which can't be done with a CNN alone. We don't include the significance of events in our study, but it will be an important step for future work using real GW detector data.
Moreover, the wavelet coefficient information produced by WDF could be used to reconstruct the waveform of a detected event, and this will be included in future work. 

We performed only a single detector search, using one current and one future third generation GW detector. In the future, our search would need to be extended to a multi detector analysis. In this case, we would run WDF on data from multiple detectors and search for coincidence between the different detectors triggers before passing the data to the CNN.

We included in our study 5 of the most modern 3D simulations of CCSNe that we used to train the CNN. We also trained the CNN on two types of simulated glitches to show how we can distinguish real events from transient detector noise events. In real GW detector data, there are a larger number of different glitch types, and the data needed to train machine learning models for the real glitches can be obtained from the citizen science project Gravity Spy \cite{2017CQGra..34f4003Z}.

We ran both a 1-D CNN on the time series data, and a 2-D CNN using spectrograms of the data. The 1-D method is less robust due to the stochastic nature of the waveforms. Although the time series features of the waveforms we used here cannot be related back to the source parameters, they are still very different to glitches, and we show that we can distinguish them from noise with a 95\% accuracy. It is likely that the 1-D search would also find GW signals that are not CCSNe, as the time series could represent any burst of GWs. However, in the 2-D case, the spectrograms used for training are directly related to emission features we only expect from a CCSN signal, and should be able to distinguish between a CCSN and some other kind of GW burst. The common features in the spectrograms improve our detection accuracy slightly. However, we must be careful during the training to include all the possible features, as we can see that the accuracy for model s25 decreases when we do not include other signals with a strong low frequency SASI component in the training. Therefore, we see the benefit in including both a 1-D and a 2-D search, as the 1-D search is more likely to detect the GWs if our predictions of the CCSN emission are incorrect, and if they are correct then the features in the 2-D spectrograms result in increased search accuracy. 
For this reason, we applied for the first time in this field the ensemble method for the 2 different classification model in the multilabel case.

We showed that the CNN can also distinguish between different CCSN waveform models. We don't have any important astrophysical differences in these models that we used here, but in future studies we could use this to distinguish between waveforms from different explosion mechanisms, or models with and without rotation.

\section*{Acknowledgments}
\label{sec:ack}
This article/publication is based upon work from COST Action CA17137, supported by COST (European Cooperation in Science and Technology).
JP is supported by the Australian Research Council (ARC) Centre of 
Excellence for Gravitational Wave Discovery (OzGrav), through project number 
CE170100004.
FM is supported by the Polish National Science Centre grant no. 2016/22/E/ST9/00037.

\section*{References}

\bibliographystyle{iopart-num}
\bibliography{bibfile}

\end{document}